\documentclass[useAMS,usenatbib]{mn2e}
\usepackage{graphics,rotating,multirow,bm,color,hyperref}
\hypersetup{
    colorlinks=true,
    linkcolor=green,
    citecolor=blue,
}

\title[Semi-analytical Ray-tracing]
  {A Semi-analytic Ray-tracing Algorithm for Weak Lensing}
\author{circulate within collaborators}
\author[Baojiu~Li, Lindsay~J.~King, Gong-Bo Zhao, Hongsheng Zhao]
  {Baojiu~Li$^{1,2,}$\thanks{E-mail: b.li@damtp.cam.ac.uk}, Lindsay~J.~King$^{2,3,}$\thanks{E-mail: ljk@ast.cam.ac.uk}, Gong-Bo Zhao$^{4,}$\thanks{E-mail: gong-bo.zhao@port.ac.uk}, Hongsheng Zhao$^{5,}$\thanks{E-mail: hz4@st-andrews.ac.uk}\\
  $^1$DAMTP, Centre for Mathematical Sciences, University of Cambridge, Wilberforce Road, Cambridge CB3 0WA, UK\\
  $^2$Kavli Institute for Cosmology Cambridge, Madingley Road, Cambridge CB3 0HA, UK\\
  $^3$Institute of Astronomy, University of Cambridge, Madingley Road, Cambridge, CB3 0HA, UK\\
  $^4$Institute of Cosmology \& Gravitation, University of Portsmouth, Dennis Sciama Building, Portsmouth, PO1 3FX, UK\\
  $^5$SUPA, University of St Andrews, North Haugh, Fife, KY16 9SS, UK}
\pagerange{\pageref{firstpage}--\pageref{lastpage}} \pubyear{2010}

\def\LaTeX{L\kern-.36em\raise.3ex\hbox{a}\kern-.15em
    T\kern-.1667em\lower.7ex\hbox{E}\kern-.125emX}

\def\ie{{\frenchspacing\it i.e.}}
\def\eg{{\frenchspacing\it e.g.}}

\newcommand{\tc}[1]{\textcolor{black}{#1}}

\begin{document}

\label{firstpage}

\maketitle

\begin{abstract}
We propose a new ray-tracing algorithm to measure the weak lensing
shear and convergence fields directly from $N$-body simulations.
We calculate the deflection of the light rays lensed by the 3-D
mass density field or gravitational potential along the line of sight
on a grid-by-grid basis, rather than using the projected 2-D lens
planes. Our algorithm uses simple analytic formulae instead of
numerical integrations in the computation of the projected density
field along the line of sight, and so is computationally efficient, accurate 
and straightforward to implement. This will prove valuable in the 
interpretation of data from the next generation of surveys that will image
many thousands of square degrees of sky.
\end{abstract}


\section{Introduction}

Weak gravitational lensing (WL) is a promising tool to map the
matter distribution in the Universe and constrain cosmological
models, using the statistical quantities primarily constructed out
of the observed correlations in the distorted images of distant
source galaxies. In 2000, four teams announced the first
observational detections of cosmic shear \citep{ba00,
ka00,vW00,wi00,ma01}. Since then improved observational results have 
been published \citep{hoekstra2006, fu2008, schrabback2010}, and it has been extensively used to
investigate key cosmological parameters such as the matter density
parameter $\Omega_{\rm m}$, and the normalisation of the matter
power spectrum $\sigma_{8}$ as well as for constraining neutrino
mass \citep{teretal2009}. Much theoretical progress has also been
made in assessing the utility of cosmic shear in, for example,
estimating the equation of state of dark energy $w$
\citep{brikin07,li2009,cpz2009}, as well as its role in testing
theories of modified gravity
\citep{schmidt08,zpsz1,zpsz2,zgp,song} and constraining
quintessence dark energy \citep{ck2010}.

On linear scales, one can use linear perturbation theory to
calculate the WL observables for a given cosmology, such as the
shear power spectrum or the aperture mass statistic, and compare
these predictions to observational data to constrain the model
parameters. However, the observables on nonlinear scales, which
cannot be predicted theoretically without the help of $N$-body
simulations, can also provide valuable information to prove, or
falsify cosmological models. Making such predictions using
$N$-body simulations becomes increasingly important as we move
into a new era in weak lensing using large observational surveys.
The next generation of cosmic shear surveys, \eg, the Dark Energy
Survey (DES; www.darkenergysurvey.org) will be more than an order 
of magnitude larger in
area than any survey to date, covering thousands of square
degrees, and using several filters that allow photometric redshift
estimates for the source galaxies to be derived. These surveys
have the potential to map dark matter in 3-D at unprecedented
precision, testing our structure formation paradigm and
cosmological model.

To obtain the statistics for WL from the outputs of $N$-body
simulations, one needs to construct numerous virtual light rays
propagating from the source to the observer. By tracing these
light rays along the lines of sight (l.o.s.), one could in
principle calculate how much the original source image is
distorted, and magnified.

Conventional ray-tracing algorithms generally
project the matter distribution along the paths of light rays onto
a series of lens-planes, and use the discrete lensing
approximation to compute the total deflection of the light rays on
their way to the observer  \citep{jsw,hhws}. The lens planes could be set up either
by handling the simulation outputs after the $N$-body simulation
is completed or by recording corresponding light cones on-the-fly
\citep{hsseev} and projecting later. Although this algorithm is
the most frequently used in the literature, it requires a large
amount of data, such as particle positions, to be stored, and this
would be difficult for simulations with very high mass resolution
or very big box sizes, which are increasingly more common today.
Furthermore, projecting particles onto a number ($\sim20-30$) of
lens planes will inevitably erase the detailed matter distribution
along the lines of sight and oversimplify the time evolution of
the large scale structure.

One can also perform the lensing computation during the $N$-body
simulation process to obtain the projected (surface) density
and/or convergence field directly \citep{wh2000}. This method
avoids the expensive storage of dump data at numerous redshifts
and allows the detailed matter distribution to be probed. However,
it does involve numerical integrations in the calculation of the
projected density field and therefore certain overheads, because
in order to make the integrals accurate one has to sample the
density field rather densely.

Motivated by the promise of cosmic shear surveys, and the need to
make predictions of observables on nonlinear scales using
cosmological simulations, in this work we introduce a new
algorithm to preform ray-tracing on the fly, which is based on
that of \citet{wh2000}. We calculate the deflection of a light ray
as it goes through the $N$-body simulation grids using the 3-D
density field inside the grids, instead of using the 
density field projected onto discrete 2-D lensing planes. Furthermore, the
numerical integration is replaced by some exact analytic formulae,
which could greatly simplify the computation. We will show our
result in comparison with the fitting formula, and discuss how our
algorithm can be applied to particle or potential outputs recorded
in large simulations, and how we can go beyond the Born
approximation and include the lens-lens coupling effect.

This paper is organised as follows. We will introduce our
algorithm in the next section, describe our simulation and present
the results in Sect.~3, and close with a section of discussion and
conclusion. Although we do not include lens-lens coupling and
corrections to the Born approximation in our simulations, we will
outline in Appendix~\ref{appen:A} how these can be done. For
simplicity, we shall consider a spatially flat universe throughout
this work, but the generalisation to non-flat geometries is
straightforward. We shall use ``grid" and ``grid cell"
interchangeably to stand for the smallest unit of the mesh in the
particle-mesh $N$-body simulations.

\section{Methodology}

In this section, we will first briefly review the traditional
`plane-by-plane' ray-tracing algorithm, and then detail our
improved `grid-by-grid' prescription.

\subsection{Conventional Ray-tracing Algorithm}

\label{subsect:oldmethod}

We work in the weak-lensing regime, meaning that the light rays
can be well approximated as straight lines \citep{mellier1999,
bs2001}. The metric element is given by
\begin{eqnarray}
ds^2 =
a^2\left[(1+2\Phi)d\tau^2-(1-2\Phi)\mathbf{dx}\cdot\mathbf{dx}\right]
\end{eqnarray}
where $a$ is the scale factor normalised so that $a=1$ today,
$\tau$ is the conformal time, $\Phi$ is the gravitational
potential and $\mathbf{x}$ the comoving coordinate. We use units
such that $c=1$.

Then the change of the photon's angular direction as it propagates
back in time is \citep{lc2009}
\begin{eqnarray}\label{eq:xi}
\vec{\xi}(\chi_s) - \vec{\xi}_0 =
-2\int^{\chi_s}_0\frac{\chi_s-\chi}{\chi\chi_s}\vec{\nabla}_{\vec{\xi}}\Phi
d\chi
\end{eqnarray}
in which $\chi$ is the comoving angular diameter distance,
$\vec{\xi}$ is the angular position perpendicular to the l.o.s.,
$\vec{\xi}_0\equiv\vec{\xi}(\chi=0)$, $\nabla_{\vec{\xi}}$ denotes
the covariant derivative on the sphere with respect to $\vec{\xi}$
and $\Phi=\Phi\left(\chi,\vec{\xi}\right)$ the gravitational
potential along the l.o.s.. The $2\times2$ distortion matrix is
given by
$A_{ij}\equiv\nabla_i\xi_j\equiv\nabla_{\xi_{0i}}\xi_j(\chi)$,
where $\xi_{0i}$ is the $i$-th component of $\vec{\xi}_0$, and is
equal to
\begin{eqnarray}\label{eq:distortmat}
A_{ij} &\equiv&
-2\int^{\chi_s}_0\frac{g(\chi,\chi_s)}{\chi^2}\nabla_{\xi_{0i}}\nabla_{\xi_j}\Phi\left(\chi,\vec{\xi}\right)
d\chi + \delta_{ij}\nonumber\\
&\approx&
-2\int^{\chi_s}_0\frac{g(\chi,\chi_s)}{\chi^2}\nabla_{\xi_{i}}\nabla_{\xi_j}\Phi\left(\chi,\vec{\xi}\right)
d\chi + \delta_{ij}
\end{eqnarray}
with $i,j=1,2$ running over the two components of $\vec{\xi}$, and
\begin{eqnarray}
g(\chi,\chi_s) \equiv \frac{(\chi_s-\chi)\chi}{\chi_s}.
\end{eqnarray}
Note that to obtain Eq.~(\ref{eq:distortmat}) we have made the
approximation $\nabla_{\xi_{0i}}\approx\nabla_{\xi_{i}}$, which
means that lens-lens coupling is ignored. We shall discuss how to
go beyond this approximation in Appendix~\ref{appen:A}.

This matrix is related to the convergence $\kappa$ and shear components
$\gamma_{1,2}$ by
\begin{eqnarray}\label{eq:distortmat2}
\mathbf{A} = \left(\begin{array}{cc}
1-\kappa-\gamma_1 & -\gamma_2-\omega \\
-\gamma_2+\omega & 1-\kappa+\gamma_1 \\
\end{array}\right)
\end{eqnarray}
where $\omega$ stands for the rotation, and
$\gamma=\left(\gamma_1^2+\gamma_2^2\right)^{1/2}$ the shear
magnitude. In the weak-lensing approximation, once the convergence
is obtained, the shear is determined as well, therefore in
practice we only need to compute $\kappa$,

\begin{eqnarray}\label{eq:kappa1}
\kappa = 1-\left(A^1_{\ 1}+A^2_{\ 2}\right)/2 = \int^{\chi_s}_0
g(\chi,\chi_s)\nabla_{\xi}^2\Phi d\chi.
\end{eqnarray}
Under the Limber approximation \citep{wh2000}, the two-dimensional
Laplacian in Eq.~(\ref{eq:kappa1}) can be replaced with the
three-dimensional Laplacian, because the component of the latter
parallel to the l.o.s. is negligible on small angular scales
\citep{jsw}. Then, using the Poisson equation
\begin{eqnarray}
\nabla^2\Phi = \frac{3}{2}\Omega_m H_0^2\frac{\delta}{a}
\end{eqnarray}
where $\delta$ is the matter overdensity and $a$ the scale factor, we can rewrite Eq.~(\ref{eq:kappa1}) as
\begin{eqnarray}\label{eq:kappa2}
\kappa(\vec{\xi}) = \frac{3}{2}\Omega_mH_0^2\int^{\chi_s}_0
g(\chi,\chi_s)\frac{\delta(\mathbf{x})}{a}d\chi
\end{eqnarray}
in which we have written explicitly the $x_{\perp}$-dependence of
$\kappa$ (the $\chi$-dependence is integrated out in the
projecting process). Eq.~(\ref{eq:kappa2}) is the starting point
of most ray-tracing simulations.

The most commonly-used ray-tracing method is the discrete lensing
approximation. In this approach, the density field is projected
onto a number of lensing planes (usually $\sim20-30$), and the
light rays are treated as if they were deflected only by these
plane lenses. Correspondingly, the term $g(\chi,\chi_s)$ in
Eq.~(\ref{eq:kappa2}) is evaluated only at the positions of these
planes.

The method of \citet{wh2000} incorporates the integration in
Eq.~(\ref{eq:kappa2}) directly into their $N$-body simulation
code, and performs the integral at every time-step. To realise
this, $N_{\rm los}$ straight l.o.s. are generated to be traced.
The rays have specified origin (the observer at redshift 0),
opening (\eg, $3^{\circ}\times3^{\circ}$) and orientation. As the
$N$-body simulation process evolves to the source redshift $z_s$,
the convergence is computed along each line of sight using
Eq.~(\ref{eq:kappa1}) or Eq.~(\ref{eq:kappa2}). The l.o.s.
integration is then carried out numerically for each time step,
during which the photon travels from  $\chi_i$ to $\chi_f$, where
the subscripts $i$ and $f$ literally stand for \emph{initial} and
\emph{final} respectively, and hence they are used as the
integration boundaries. The integrand $g\nabla^2_\xi\Phi/\chi^2$
in Eq.~(\ref{eq:kappa1}) or $g\delta(\mathbf{x})$ in
Eq.~(\ref{eq:kappa2}) is considered to be constant during each
time-step, and the integral is approximated by summing over all
the time steps. The time sampling has to be sufficiently fine so
as to guarantee the required numerical accuracy.

One advantage of this algorithm is that $\kappa$ is computed
step-by-step on the fly, so one can avoid the expensive disk
storage required for storing particle dumps and the time-consuming
postprocessing analysis. Moreover, in this approach, there is no
difficulty to make ultra-fine time sampling -- the number of time
slices can be as many as the number of time steps for the
simulation (after $z_s$), which is a mission impossible for the
postprocessing approach -- making the result more accurate than
the postprocessing approach.

However, one does have to carry out the numerical integration in
Eq.~(\ref{eq:kappa1}) or Eq.~(\ref{eq:kappa2}), and to make the
result accurate one has to sample the value of the integrand very
densely (\eg, $\sim100$ sampling points are dynamically chosen for
each time step), which might cause certain overheads when a large
number of light rays are traced and ultra-fine time-stepping is
used.

\subsection{Improved Ray-tracing Algorithm}

In this work, we propose an improved ray-tracing algorithm by
computing the convergence, shear and projected density fields on
the very grid cells on which the $N$-body simulation is performed.
In our grid-by-grid approach, the l.o.s integration can be carried
out \emph{analytically}, making the computation more efficient and
accurate. Also, the light rays are deflected by the detailed
matter distribution exactly as seen in the $N$-body simulations,
making the ray-tracing and $N$-body simulations consistent with
each other. A detailed derivation of the relevant formulae is
given in Sections~\ref{subsect: method1} \& \ref{subsect:
method2}, and the basic idea is as follows. Take
Eq.~(\ref{eq:kappa2}) as an example, the integrand is
$g\delta(\mathbf{x})$. Since our particle-mesh ({\tt PM}) code
automatically computes $\delta(\mathbf{x})$ on the regular mesh,
the value of $\delta(\mathbf{x})$ at any point can be obtained by
interpolation, and in particular we can compute the value along
the line of sight as a function of the comoving distance $\chi$,
and the values of $\delta(\mathbf{x})$ at the vertices of the grid
containing the said point \footnote{Because the line of sight is
by approximation a straight line, once the comoving distance
$\chi$ to a point is known, the corresponding $x,y,z$-coordinates
of that point can be expressed in terms of $\chi$ and orientation
angles, which are fixed when the l.o.s. are assumed to be
straight.}. Note that the vertices themselves are regular grid
points, and the values of $\delta(\mathbf{x})$ on the vertices are
known. Using certain interpolation schemes, trilinear, for
example, $\delta(\mathbf{x})$ can be expressed as a polynomial of
$\chi$, thus the integral can be carried out analytically.
Therefore, no numerical integration is needed to compute $\kappa$.
Similarly, our algorithm can also be used to compute the integral
of Eq.~(\ref{eq:kappa1}) analytically, as detailed in
Section~\ref{subsect: method2}.

Note that when the algorithm is applied to the density field
$\delta(\mathbf{x})$, there is some subtlety, and this will be
clarified in Sect.~\ref{subsect: method1}.

\begin{figure*}
\includegraphics[width=160mm]{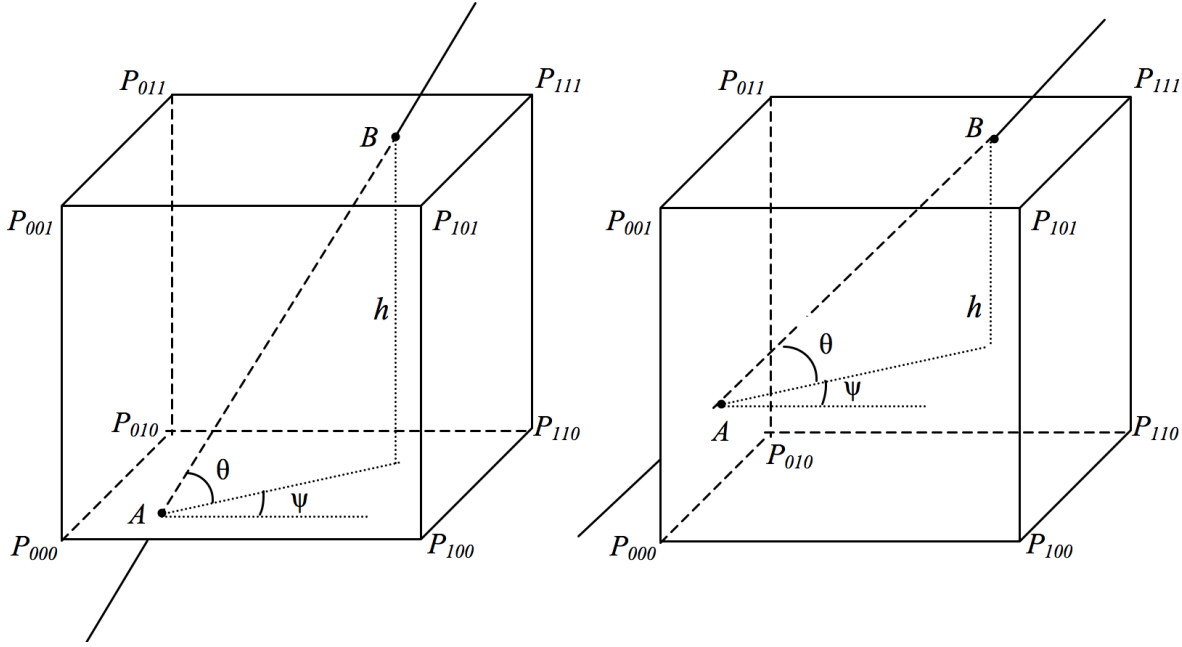}
\caption{Two examples of the line-of-sight (lines) crossing a cubic
cell of the simulation box at points $A$ and $B$. The segment $AB$
lies inside the cube. $P_{xyz}$ ($x,y,z=0,1$) are the eight vertexes
of the cube. Projecting $AB$ into the plane perpendicular to $z$-direction
and passing point $A$, then $\psi$ is the angle between the projection
and $x$-direction, and $\theta$ is the angle between $AB$ and that
plane. For the given line-of-sight and cube, $A, B, \theta, \psi$ are
known or can be computed easily, and we also know the density values at
the eight vertexes; we then want to integrate the density field along
$AB$ (or part of it).}
\label{fig:cube}
\end{figure*}

\subsubsection{Method A}

\label{subsect: method1}

To compute the projected density field, we need to integrate along
the l.o.s., and in practice this integration can be carried out
progressively along the segments of lines of sight within
individual cubic grid cells. The reason for such a prescription
will become clear soon. Throughout this subsection we will use $z$
to denote the coordinate rather than redshift.

Fig.~\ref{fig:cube} shows two examples of such configurations, in
which the part $AB$ of a line of sight lies in a grid (see the
figure caption for more information). The density value at a given
point on $AB$ could be computed using trilinear interpolation, as
long as we know the values at the vertices, which are denoted by
$\rho_{xyz}$ ($x,y,z=0,1$). To be more explicit, let us define
\begin{eqnarray}\label{eq:coeff_cs}
c_0 &\equiv& \rho_{000},\nonumber\\
c_1 &\equiv& \rho_{100}-\rho_{000},\nonumber\\
c_2 &\equiv& \rho_{010}-\rho_{000},\nonumber\\
c_3 &\equiv& \rho_{001}-\rho_{000},\nonumber\\
c_4 &\equiv& \rho_{110}-\rho_{010}-\rho_{100}+\rho_{000},\nonumber\\
c_5 &\equiv& \rho_{011}-\rho_{001}-\rho_{010}+\rho_{000},\nonumber\\
c_6 &\equiv& \rho_{101}-\rho_{001}-\rho_{100}+\rho_{000},\nonumber\\
c_7 &\equiv& \rho_{111}-\rho_{011}-\rho_{101}-\rho_{110}\nonumber\\
&&+\rho_{100}+\rho_{001}+\rho_{010}-\rho_{000}.
\end{eqnarray}
Suppose the point on $AB$ we are considering has coordinate
$(x,y,z)$, then the density value is given by
\begin{eqnarray}\label{eq:rho_xyz1}
\rho(x,y,z) &=& c_0 + c_1\Delta x + c_2\Delta y + c_3\Delta z + c_4\Delta x\Delta y\nonumber\\
&& + c_5\Delta y\Delta z + c_6\Delta x\Delta z + c_7\Delta x\Delta y\Delta z,
\end{eqnarray}
where
\begin{eqnarray}
\Delta x &\equiv& \frac{1}{L}\left(x-x_0\right)\ =\ \frac{1}{L}\left[a+(\chi-\chi_A)\cos\theta\cos\psi\right],\nonumber\\
\Delta y &\equiv& \frac{1}{L}\left(y-y_0\right)\ =\ \frac{1}{L}\left[b+(\chi-\chi_A)\cos\theta\sin\psi\right],\nonumber\\
\Delta z &\equiv& \frac{1}{L}\left(z-z_0\right)\ =\
\frac{1}{L}\left[c+(\chi-\chi_A)\sin\theta\right]
\end{eqnarray}
where $L$ denotes the size of the cubic cell, $(x_0,y_0,z_0)$ is
the coordinate of vertex $P_{000}$, $\chi_A$ is the $\chi$ value
at point $A$, and $a, b, c$ are the coordinates of point $A$
relative to $P_{000}$.

Because we express $\rho(x,y,z)$ in terms of $\chi$ only, the line
integral along $AB$ could be rewritten as an integral over $\chi$,
and we have
\begin{eqnarray}\label{eq:los_int}
&&\frac{1}{\chi_s}\int^{\chi_{u}}_{\chi_l}\chi(\chi_s-\chi)\rho(x,y,z)d\chi\nonumber\\
&=& \frac{1}{\chi_s}\sum_{N=1}^4d_N\int^{\chi_u}_{\chi_l}\chi(\chi_s-\chi)(\chi-\chi_A)^{N-1}d\chi\nonumber\\
&=& \frac{1}{\chi_s}\sum_{N=1}^4d_N\int^{\tilde{\chi}_u}_{\tilde{\chi}_l}\left(\tilde{\chi}+\chi_A\right)\left(\chi_s-\chi_A-\tilde{\chi}\right)\tilde{\chi}^{N-1}d\tilde{\chi}\nonumber\\
&=& \sum_{N=1}^4\frac{d_N}{N}\chi_A\left(1-\frac{\chi_A}{\chi_s}\right)\left(\tilde{\chi}^N_u-\tilde{\chi}^N_l\right)\nonumber\\
&&+\sum_{N=1}^4\frac{d_N}{N+1}\left(1-2\frac{\chi_A}{\chi_s}\right)\left(\tilde{\chi}^{N+1}_u-\tilde{\chi}^{N+1}_l\right)\nonumber\\
&&-\sum_{N=1}^4\frac{d_N}{N+2}\frac{1}{\chi_s}\left(\tilde{\chi}^{N+2}_u-\tilde{\chi}^{N+2}_l\right),
\end{eqnarray}
in which $\chi_l\geq \chi_A$ and $\chi_u\leq \chi_B$\footnote{Note
that $A$ and $B$ are the intersections between the l.o.s.~and the
grid cell, and not necessarily the two ends of the l.o.s.~in one
time step. But the integration is carried out for each time step,
and so we do not always have $\chi_l=\chi_A$ and $\chi_u=\chi_B$.}
are the lower and upper limit of the integral respectively,
$\tilde{\chi}\equiv\chi-\chi_A$,
$\tilde{\chi}_u\equiv\chi_u-\chi_A$,
$\tilde{\chi}_l\equiv\chi_l-\chi_A$ and we have also defined
\begin{eqnarray}\label{eq:ds}
d_1 &\equiv& c_0 + \frac{1}{L}\left(ac_1+bc_2+cc_3\right)\nonumber\\
&&+\frac{1}{L^2}\left(abc_4+bcc_5+acc_6\right)+\frac{1}{L^3}abcc_7,\nonumber\\
d_2 &\equiv& \frac{1}{L}\cos\theta\cos\psi c_1 + \frac{1}{L}\cos\theta\sin\psi c_2 + \frac{1}{L}\sin\theta c_3\nonumber\\
&&+\frac{1}{L^2}\cos\theta\sin\psi\left(ac_4+cc_5\right)\nonumber\\
&&+\frac{1}{L^2}\cos\theta\cos\psi\left(bc_4+cc_6\right)\nonumber\\
&&+\frac{1}{L^2}\sin\theta\left(bc_5+ac_6\right)+\frac{1}{L^3}\sin\theta abc_7\nonumber\\
&&+\frac{1}{L^3}\cos\theta\sin\psi acc_7+\frac{1}{L^3}\cos\theta\cos\psi bcc_7,\nonumber\\
d_3 &\equiv& \frac{1}{L^2}\cos^2\theta\sin\psi\cos\psi c_4+\frac{1}{L^2}\sin\theta\cos\theta\sin\psi c_5\nonumber\\
&&+\frac{1}{L^2}\sin\theta\cos\theta\cos\psi c_6+\frac{1}{L^3}\sin\theta\cos\theta\sin\psi ac_7\nonumber\\
&&+\frac{1}{L^3}\sin\theta\cos\theta\cos\psi bc_7\nonumber\\
&&+\frac{1}{L^3}\cos^2\theta\sin\psi\cos\psi cc_7,\nonumber\\
d_4 &\equiv& \frac{1}{L^3}\sin\theta\cos^2\theta\sin\psi\cos\psi
c_7.
\end{eqnarray}
Note that, by writing the result in the above form, we have
separated the treatments for four types of variables:
\begin{enumerate}
\item $a, b, c, \chi_A, \tilde{\chi}_{u,l}$: $a, b, c, \chi_A$ are
determined by the direction of the light ray and the specific grid cell under
consideration, and $\tilde{\chi}_{u,l}$ depend only on the
considered time step and $\chi_A$. Note that $a,b,c$ must be determined 
carefully, and for each grid cell at least one of them vanishes, but exactly 
which of them vanishes varies from ray to ray and from grid cell to grid cell;  
\item $\theta, \psi$: these
specify the direction of the light ray, and terms involving them
only need to be computed once, \ie, at the beginning of the
simulation, for a given line of sight; \item $c_{0-7}$ -- these
are determined by the values of $\rho$ at the vertices of a grid,
and must be evaluated for each grid that the light ray passes
through; \item $L, \chi_s$: these are constants for a given
simulation.
\end{enumerate}
Therefore once $\rho_{xyz}$ is known, the integral can be
performed analytically without much computational effort. This is
not unexpected, because once the density is known at the vertices
of the grid, we should know the density at any point inside the
grid using interpolation, and no more information is needed to
carry out the integral. If we consider a different grid, a
different set of $\rho_{xyz}$ needs to be used, and this is why
our algorithm is based on the individual grids.

There are two technical points which need to be noted. First, in
Eq.~(\ref{eq:los_int}) $\rho$ should be replaced by $\rho/a$ in
practice. It is true that $a$ could be expressed as a function of
$\chi$ as well once the background cosmology is specified, but
this will lead to more complicated expressions. Therefore in our
simulations we simply take $a$ to be constant during each time
step. This is certainly only an approximation, but we should note
that $a$ is considered as constant during each time step in the
$N$-body simulations anyway. Indeed, as we see in
Section~\ref{subsect: method2}, the factor $\frac{1}{a}$ does
not appear if we use $\nabla^2_{\xi}\Phi$ instead of $\rho$ in the
integral\footnote{This just reflects the fact that during each
time step of the $N$-body simulation, the $1/a$ factor in the
Poisson equation is treated as constant. The nature of numerical
simulation (discreteness in time) dictates that we cannot do
better save decreasing the length of time-steps, which we cannot
always keep doing in reality.}.

Second, as has been mentioned by various papers ({\it
e.g.} \citet{jsw, wh2000}), the use of the three dimensional Laplacian
[Eq.~(\ref{eq:kappa2})] instead of the two dimensional one
[Eq.~(\ref{eq:kappa1})] is at best an approximation. We have to
test the validity of this approximation. In fact, as we show
below, the error caused by this approximation is actually not
negligible. To see this, recall that
\begin{eqnarray}\label{eq:int_by_part}
\kappa &=& \int^{\chi_s}_0g\left(\nabla^2-\nabla^2_\chi\right)\Phi d\chi\nonumber\\
&=& \frac{3}{2}\Omega_mH_0^2\int^{\chi_s}_0g\frac{\delta}{a}d\chi - \left[g\nabla_\chi\Phi\right]^{\chi_s}_0\nonumber\\
&&+\int^{\chi_s}_0g\dot{\nabla}_\chi\Phi d\chi +\int^{\chi_s}_0g'\nabla_\chi\Phi d\chi
\end{eqnarray}
in which a prime (overdot) denotes the $\chi$ (time) derivative,
and the last three terms come from the treatment of
$\nabla^2_\chi\Phi$, including integration by parts. The common
argument is that the second term actually vanishes as $g=0$ and
$\nabla_\chi\Phi<\infty$ at $\chi=\chi_s$ and $\chi=0$, and the
last two terms are negligible. This is true in the ideal case, but
while our algorithm [and that of \citet{wh2000}] is applied the
second term is no longer zero because of the following reasons:
\begin{enumerate}
\item It is unrealistic to make the simulation boxes big enough to
contain the whole light cone, and in practice people tile
different simulations to form a complete light cone. Unless a
periodic tiling of the same box is adopted, we expect the matter
distribution and thus the potential $\Phi$ to be discontinuous at the
tiling boundaries. As a result the second term in
Eq.~(\ref{eq:int_by_part}) should read
\begin{eqnarray}
\left[g\nabla_\chi\Phi\right]^{\chi_s}_0 = \left[g\nabla_\chi\Phi\right]^{\chi_s}_{\chi_{l1}} + \left[g\nabla_\chi\Phi\right]^{\chi_{u2}}_{\chi_{l2}} + \cdots + \left[g\nabla_\chi\Phi\right]^{\chi_{uN}}_0\nonumber
\end{eqnarray}
in which $\chi_{u,l}$ correspond to the values of $\chi$ when the
light ray {goes through} a given box, which is labelled as $1, 2,
\cdots, N$. If the matter distribution is smooth at the boundaries
of the boxes, then
$\nabla_\chi\Phi(\chi=\chi_{l1})=\nabla_\chi\Phi(\chi=\chi_{u2})$
and so on, so all terms cancel. However, if the matter
distribution is not smooth, as is the case for many tiling
treatments, then such cancelling will not happen and
$\left[g\nabla_\chi\Phi\right]^{\chi_s}_0$ will turn out to be
nonzero in the numerical calculation although it should be zero in
theory. \item Using the same argument as above, we could find that
this discontinuity problem appears not only on the boundaries of
the tiled simulation boxes, but also at {\it each time-step} in
the simulations and {\it each time} when the light ray passes
through a grid of the simulation box. For the former case, suppose
that during one time step the l.o.s.~ends at point $C$, then $C$
is also the point where this l.o.s.~starts during the next time
step. However, the values of $\nabla_{\chi}\Phi$ at point $C$ are
generally different in the two time steps because particles have
been advanced, and so a discontinuity appears. For the latter
case, our piecewise l.o.s.~integral and the interpolation scheme
dictate that the values of $\nabla_\chi\Phi$ at a point $D$ on the
interface of two neighbouring grids could depend on which grid is
supposed to contain point $D$ (remember the interpolation scheme
uses the values of $\nabla_\chi\Phi$ at the vertices of the {\it
containing} cell), and naturally a discontinuity in
$\nabla_\chi\Phi$ appears at the interface of the two grids. Note
that these discontinuities are inevitable due to the nature of
numerical simulation (the discreteness in time), and decreasing
the grid size or the length of time steps does not help because
then such discontinuities will only appear more
frequently\footnote{Interestingly, the discrete lens-plane
approximation does not have this problem (as long as simulation
boxes are tiled periodically so that matter distribution is smooth
on the tiling boundaries), because it does not treat the
l.o.s.~integral on a grid-by-grid basis.}.
\end{enumerate}

The way to tackle these problems is as follows: we know that
$\left[g\nabla_\chi\Phi\right]^{\chi_s}_0$ vanishes rigorously in
principle but is nonzero because of the nature of the simulation;
meanwhile, the same discontinuity problem also appears when
calculating the first quantity on the right-hand side of
Eq.~(\ref{eq:int_by_part}). The errors in the numerical values for
these two quantities are caused by the same discontinuity and
could cancel each other. The exact value of this error can be
obtained by computing $\left[g\nabla_\chi\Phi\right]$, because
this quantity is zero in theory and its nonzero value is
completely the error. In our simulations, we compute
$\left[g\nabla_\chi\Phi\right]$ explicitly whenever the light ray
passes a grid, and subtract it according to
Eq.~(\ref{eq:int_by_part}): this way we can eliminate the error in
the integration of $g\delta/a$ due to the discontinuities.

As for the third and fourth terms in Eq.~(\ref{eq:int_by_part}),
the third term is nonzero but small in reality, but in our
simulations it vanishes because $\Phi$ is assumed to be constant
during any given time-step. This will cause certain unavoidable
errors, that we anyway expect to be small. The fourth term has as
small a contribution, but fortunately we can perform the integral
exactly and analytically as we have done for the first term in
Eq.~(\ref{eq:int_by_part}).

We have run several tests to check the accuracy of the
approximations, and found the following:
\begin{enumerate}
\item If we simply replace the two dimensional Laplacian in
Eq.~(\ref{eq:kappa1}) with a three dimensional one, as in
Eq.~(\ref{eq:kappa2}), then the difference is of order 10\% and
even much larger for the rays for which $|\kappa|$ is small. Note
that Eq.~(\ref{eq:kappa1}) can be evaluated exactly as will be
described in Section~\ref{subsect: method2}. \item If we
explicitly calculate the term
$\left[g\nabla_\chi\Phi\right]^{\chi_s}_0$ for each cell crossed
by a ray, and subtract it according to Eq.~(\ref{eq:int_by_part}),
the difference between Eqs.~(\ref{eq:kappa1}, \ref{eq:kappa2}) is
brought down to the level of 1-2\%. \item If we further include
the contribution from the fourth term of
Eq.~(\ref{eq:int_by_part}), the difference will fall well within
the percent level.
\end{enumerate}

\subsubsection{Method B}

\label{subsect: method2}

The method described in Section~\ref{subsect: method1} is only
applicable to Eq.~(\ref{eq:kappa2}), while there are also
motivations for us to consider Eq.~(\ref{eq:kappa1}). For example,
the use of the three-dimensional Laplacian instead of the
two-dimensional Laplacian in Eq.~(\ref{eq:kappa2}) is at best an
approximation and only works well on small angular scales. This is
even worse in the discrete lensing approximation, because the
photons of equal distance from the observer are certainly not in a
plane but on a spherical shell, and this has motivated more
accurate treatments such as the prescription proposed by
\citet{vw2003}. As another example, within the current framework
the shear is not computed directly but from its relation with
$\kappa$. There is certainly no problem with this, but it will be
even better if we can compute $\gamma_{1,2}$ directly and compare
with the results obtained from $\kappa$.

Our generalised treatment here is quite simple,  taking advantage
of the fact that the particle-mesh codes also give us the values
of $\Phi(\mathbf{x})$ and (if necessary) $\nabla_i\nabla_j\Phi$ at
the regular grid points. For simplicity, let us assume that (1)
the central line of sight is parallel to the $x$-axis, and (2) the
opening of the lines-of-sight bundle is a square with its sides
parallel to $y,z$-axes respectively. In the two-dimensional plane
perpendicular to the line of sight, the $i=1,2$ directions are set
to be longitude and latitude respectively. We also define
\begin{eqnarray}
\mu &\equiv& \nabla_x\nabla_x\Phi,\nonumber\\
\nu &\equiv& \nabla_y\nabla_y\Phi,\nonumber\\
\eta &\equiv& \nabla_z\nabla_z\Phi,\nonumber\\
\zeta &\equiv& \nabla_x\nabla_y\Phi\ =\ \nabla_y\nabla_x\Phi,\nonumber\\
\omega &\equiv& \nabla_y\nabla_z\Phi\ =\ \nabla_z\nabla_y\Phi,\nonumber\\
\varpi &\equiv& \nabla_x\nabla_z\Phi\ =\ \nabla_z\nabla_x\Phi
\end{eqnarray}
to lighten the notation. Then, given the values of $\mu, \nu,
\cdots$ at the vertices of a grid, their values at any point
inside that grid can be obtained using trilinear interpolation
just as we have done for $\rho$ in Section~\ref{subsect: method1}.

Now, for the configuration depicted in Fig.~\ref{fig:cube} we
have, after some exercise of geometry,
\begin{eqnarray}\label{eq:express}
\nabla_1\nabla_1\Phi
&=& \chi^2\cos^2\theta\left(\mu\sin^2\psi+\nu\cos^2\psi-\zeta\sin2\psi\right),\nonumber\\
\nabla_2\nabla_2\Phi
&=& \chi^2\left(\mu\cos^2\psi\sin^2\theta+\nu\sin^2\psi\sin^2\theta+\eta\cos^2\theta\right)\nonumber\\
&&+\chi^2\zeta\sin2\psi\sin^2\theta-\chi^2\varpi\cos\psi\sin2\theta\nonumber\\
&&-\chi^2\omega\sin\psi\sin2\theta,\nonumber\\
\nabla_1\nabla_2\Phi
&=& \frac{1}{2}\chi^2\sin2\theta\left[\frac{1}{2}(\nu-\mu)\sin2\psi+\zeta\cos2\psi\right]\nonumber\\
&&+\chi^2\cos^2\theta\left(\varpi\sin\psi-\omega\cos\psi\right),\nonumber\\
\nabla_\chi\nabla_\chi\Phi
&=& \mu\cos^2\psi\cos^2\theta + \nu\sin^2\psi\cos^2\theta + \eta\sin^2\theta\nonumber\\
&&+\zeta\sin2\psi\cos^2\theta+\varpi\cos\psi\sin2\theta\nonumber\\
&&+\omega\sin\psi\sin2\theta.
\end{eqnarray}

Note that the above expressions are all linear in $\mu, \nu,
\cdots$, making the situation quite simple. As an example, for
$\nabla^2_\xi\Phi = \nabla^1\nabla_1\Phi+\nabla^2\nabla_2\Phi$ we
have
\begin{eqnarray}
\nabla^2_\xi\Phi &=& \left(\sin^2\psi+\cos^2\psi\sin^2\theta\right)\mu \nonumber\\
&&+\left(\cos^2\psi+\sin^2\psi\sin^2\theta\right)\nu + \eta\cos^2\theta\nonumber\\
&&-\zeta\sin2\psi\cos^2\theta-\varpi\cos\psi\sin2\theta-\omega\sin\psi\sin2\theta,\nonumber
\end{eqnarray}
as is consistent with \citet{chs2005}, and because $\psi, \theta$ are constants for a given ray
\begin{eqnarray}\label{eq:exact_kappa}
&&\int^{\chi_s}_0\frac{\chi(\chi_s-\chi)}{\chi_s}\nabla^2_\xi\Phi d\chi\nonumber\\
&=& \left(\sin^2\psi+\cos^2\psi\sin^2\theta\right)\langle\mu\rangle-\langle\zeta\rangle\sin2\psi\cos^2\theta\nonumber\\
&&+\left(\cos^2\psi+\sin^2\psi\sin^2\theta\right)\langle\nu\rangle + \langle\eta\rangle\cos^2\theta\nonumber\\
&&-\langle\varpi\rangle\cos\psi\sin2\theta-\langle\omega\rangle\sin\psi\sin2\theta,
\end{eqnarray}
where
\begin{eqnarray}
\langle\mu\rangle &\equiv& \int^{\chi_s}_0\frac{\chi(\chi_s-\chi)}{\chi_s}\mu d\chi
\end{eqnarray}
(and similarly $\langle\nu\rangle, \cdots$) are computed exactly
as in Eq.~(\ref{eq:los_int}). Note that we only need to compute
$\langle\mu\rangle, \cdots$ during the $N$-body simulations and
multiply appropriate coefficients as in Eq.~(\ref{eq:exact_kappa})
to obtain $\kappa$ finally. The components of the shear field
$(\gamma_1,\gamma_2)$ could be computed using the same formula as
Eq.~(\ref{eq:exact_kappa}), but with $\nabla_{\xi}^2\Phi$ replaced
with $\nabla^1_{\ 1}\Phi-\nabla^2_{\ 2}\Phi$ and $\nabla^1_{\
2}\Phi$ correspondingly using the expressions given in
Eq.~(\ref{eq:express}).

\begin{figure}
\includegraphics[width=85mm]{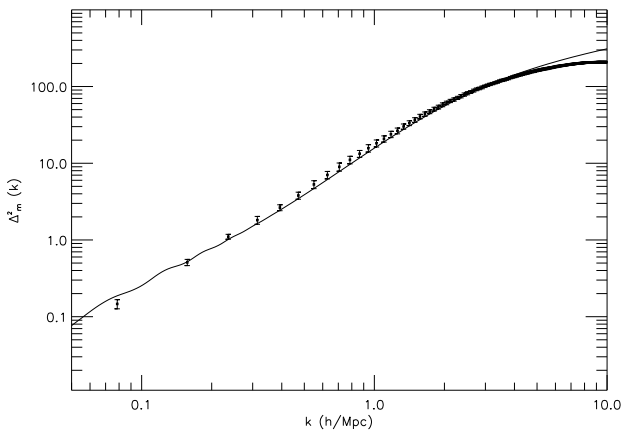}
\caption{Plotted is the $\Delta^2_m(k)\equiv
k^3P(k)/\left(2\pi^2\right)$, in which $P(k)$ is the matter power
spectrum, as a function of the wavenumber $k$ in units of
$h$~Mpc$^{-1}$. The symbols with error bars represent averaged
results at $z=0$ from 10 realisations for the $B=80h^{-1}$~Mpc
simulations. The solid curve is the corresponding result using the
\citet{setal2003} fit and the same set of cosmological
parameters.} \label{fig:power}
\end{figure}

\begin{figure*}
\includegraphics[scale=0.35]{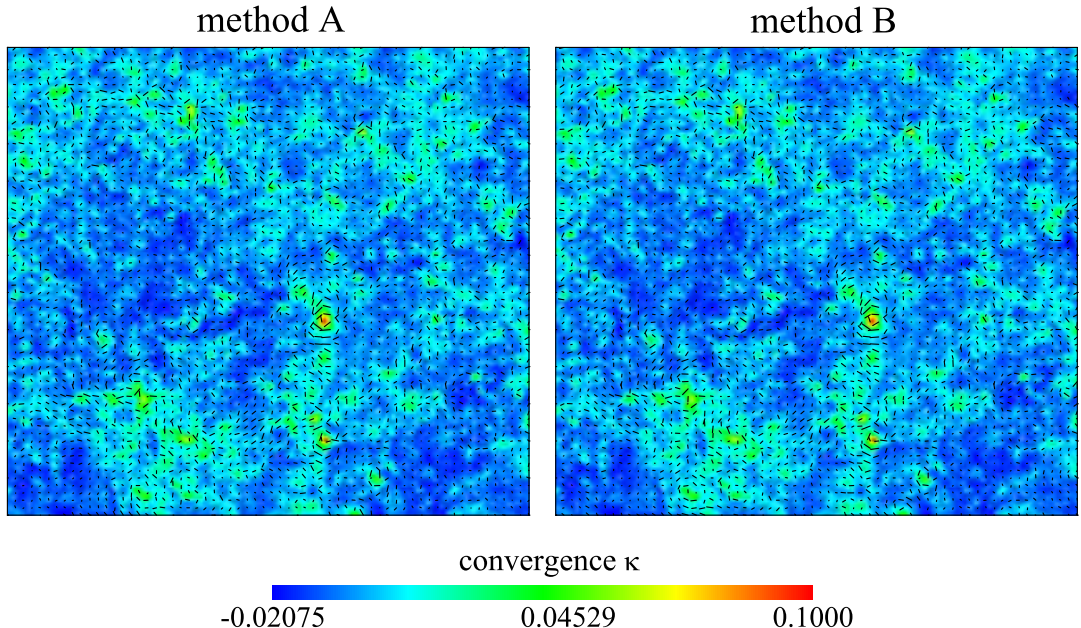}
\caption{Convergence and shear maps ($5^{\circ}\times5^{\circ}$) for one realisation from the tiling solutions. The convergence field $\kappa$ is shown as a colour-scale plot and the values are indicated by the colour bar below; the shear field $(\gamma_1,\gamma_2)$ is shown as a flow plot and superposed on the $\kappa$ field for comparison. As expected, the shear field is tangential around high-$\kappa$ regions. Note that: (1) the $\kappa$ field in the left panel is computed according to Eq.~(\ref{eq:int_by_part}) with the last three correction terms incorporated as described in Section~\ref{subsect: method1}; (2) the $\kappa$ field in the right panel is computed using Eq.~(\ref{eq:exact_kappa}); (3) the $(\gamma_1,\gamma_2)$ field in the left panel is computed using Eq.~(\ref{eq:exact_kappa}) but with $\nabla_{\perp}^2\Phi$ replaced by $\nabla_1^2\Phi-\nabla_2^2\Phi$ (for $\gamma_1$) and $\nabla_1\nabla_2\Phi$ (for $\gamma_2$); (4) the $(\gamma_1,\gamma_2)$ field in the right panel is computed indirectly by Fourier transforming $\kappa(\vec{\theta})$ to $\tilde{\kappa}(\mathbf{l})$, computing $(\tilde{\gamma}_1,\tilde{\gamma}_2)=\left(\frac{l_1^2-l_2^2}{l_1^2+l_2^2}\tilde{\kappa},\frac{2l_1l_2}{l_1^2+l_2^2}\tilde{\kappa}\right)$ and finally inverse Fourier transforming $(\tilde{\gamma}_1,\tilde{\gamma}_2)$ to $(\gamma_1,\gamma_2)$.}
\label{fig:map}
\end{figure*}

\section{$N$-body and Ray-tracing Simulations}

\label{sect:results}

To test our algorithm, we have performed a series of $N$-body
simulations for a concordance cosmology using the publicly
available code {\tt MLAPM} \citep{kgb2001}. As it is not our
intention to carry out very high-resolution simulations here, we
only use the particle-mesh part of {\tt MLAPM} so that our
simulation grid is not self-adaptively refined. We have also
developed a {\tt C} code, {\tt RATANA} (which stands for ANAlytic
RAy-Tracing), to compute the convergence and shear fields
on-the-fly as described in the above section. This section is
devoted to a summary of our results.

\subsection{Specifications for $N$-body Simulations}

We consider a concordance cosmology with cosmological parameters
$\Omega_m=0.257$, $\Omega_{\Lambda}=0.743$, $h\equiv
H_0/(100~\mathrm{km/s/Mpc})=0.719$, $n_s=0.963$ and
$\sigma_8=0.769$. The simulations start at an initial redshift
$z_i=49.0$, and initial conditions ({\it i.e.}, initial
displacements and velocities of particles) are generated using
{\tt GRAFIC} \citep{eb1995}. In this work we only consider a
source redshift $z_s=1.0$, though other values of $z_s$ or even
multiple source redshifts can easily be implemented. The
field-of-view is $5^{\circ}\times5^{\circ}$, and we trace $1024^2$
light rays.

A source at redshift $z_s=1.0$ is about $2374h^{-1}$~Mpc away from
us ($z=0$) in terms of comoving angular diameter distance, and it
is unrealistic for us to have a simulation box which is large
enough to cover the whole light-cone. In this work we adopt the
tiling scheme introduced by \citet{wh2000}. They use multiple
simulation boxes to cover the light-cone between $z=0$ and $z_s$,
and the sizes of the simulation boxes are adjusted so that smaller
boxes are used as the light rays get closer to the observer. It
has been argued that the use of multiple tiling boxes can
compensate the lack of statistical independence of fluctuations
caused by using the same simulation box repeatedly. Also the
variable box sizes mean that one can get better angular
resolutions by using smaller boxes near the observer.

Similar to \citet{wh2000}, we choose six different box-sizes and
20 tiles between $z=0$ and $z_s$, and the details are summarised
in Table~\ref{tab: tiling}. For the $N$-body simulations
(regardless of the box sizes), we use a regular mesh with
$512\times512\times512$ cubic cells. We use the triangular-shaped
cloud ({\tt TSC}) scheme to assign the matter densities in the
grid cell, and to interpolate the forces \citep{he1981, kgb2001}.
Given the matter densities in the cells, the gravitational
potential $\Phi$ is computed using fast Fourier transform ({\tt
FFT}), and the gravitational forces (first derivatives of $\Phi$)
as well as the second derivatives of $\Phi$ are then obtained by
performing finite differences. These derivatives of $\Phi$ are
subsequently utilised by {\tt RATANA} to compute the convergence
and shear fields as described in the above section.
\begin{table}
\label{tab: tiling} \caption{The tiling solution of our $N$-body
simulations.  Here $a_{\mathrm{out}}$ is the scale factor at the
time when the light rays which are traced leave a given tile, and
$B$ is the size of the simulation box in units of $h^{-1}$~Mpc.
Each simulation uses exactly 400 time steps from $z=49$ to $z=0$.
$N_{\mathrm{real}}$ is the number of realisations for each value
of box size. To obtain a tiling solution we {\it randomly} pick
out two {\it different} simulation boxes with $B=240$, two with
$B=200$, two with $B=160$, two with $B=120$, two with $B=100$ and
10 with $B=80$ -- a total of 20 simulation boxes of different
sizes.}
\begin{tabular}{@{}lccccc}
\hline\hline
$a_{\mathrm{out}}$ & $B (h^{-1}~\mathrm{Mpc})$ & $N_{\mathrm{real}}$ & $a_{\mathrm{out}}$ & $B (h^{-1}~\mathrm{Mpc})$ & $N_{\mathrm{real}}$\\
\hline
$0.527$ & $240$ & $10$ & $0.799$ & $80$ & $20$\\
$0.561$ & $240$ & $-$ & $0.819$ & $80$ & $-$\\
$0.593$ & $200$ & $10$ & $0.838$ & $80$ & $-$\\
$0.628$ & $200$ & $-$ & $0.860$ & $80$ & $-$\\
$0.657$ & $160$ & $10$ & $0.880$ & $80$ & $-$\\
$0.686$ & $160$ & $-$ & $0.902$ & $80$ & $-$\\
$0.711$ & $120$ & $10$ & $0.927$ & $80$ & $-$\\
$0.735$ & $120$ & $-$ & $0.951$ & $80$ & $-$\\
$0.757$ & $100$ & $10$ & $0.976$ & $80$ & $-$\\
$0.780$ & $100$ & $-$ & $1.000$ & $80$ & $-$\\
\hline
\end{tabular}
\end{table}

Note that unlike in many other works, we use the same grid for
both $N$-body and ray-tracing simulations. The {\tt TSC} scheme we
are using then results in some small-scale details of the matter
distribution being smoothed out, as compared to the conventional
nearest grid point ({\tt NGP}) or cloud-in-cell ({\tt CIC})
density-assignment schemes\footnote{In the {\tt TSC} scheme, the
density on a grid cell depends on the distribution of particles on
all the 26 neighbouring grid cells; in the {\tt CIC} ({\tt NGP})
scheme, it depends on the matter distribution on the 6 direct
neighbouring grid cells (the particles in that cell only).}. We
will comment on this point later.

\begin{figure}
\includegraphics[width=85mm]{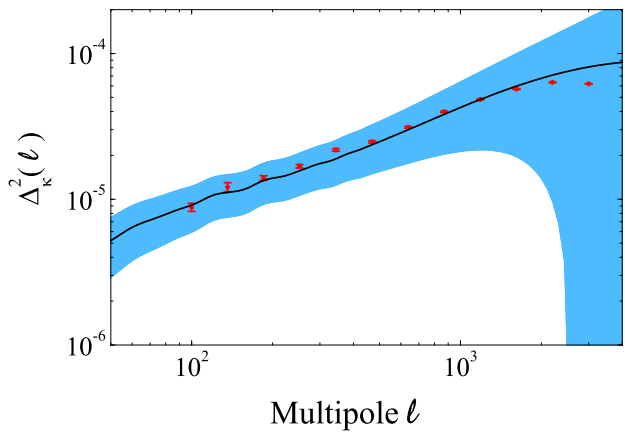}
\caption{The convergence power spectrum $\Delta^2_\kappa\equiv l^2C_l/(2\pi)$ as measured from our ray-tracing simulations (symbols with error bars). The result is obtained by averaging 120 realisations of the tiling solution. The solid curve is again obtained using the \citet{setal2003} fit of the matter power spectrum \citep{k1992, js1997}, and the filled band illustrates the expected observational uncertainty from DES.}
\label{fig:pkappa}
\end{figure}

\subsection{Numerical Results}

In this subsection we summarise the numerical results from our
$N$-body and ray-tracing simulations.

In Fig.~\ref{fig:power} we compare the matter power spectra (or
equivalently $\Delta^2_m(k)$ defined in the figure caption)
computed from our $N$-body simulations (box size 80$h^{-1}$~Mpc)
to the prediction of the analytic fitting formula of
\citet{setal2003}. We can see a good agreement, except in the
range of $0.5h~\mathrm{Mpc}^{-1} \leq k \leq
2.0h~\mathrm{Mpc}^{-1}$ where the $N$-body simulations predict a
slightly higher power. However, the agreement becomes poor for
$k>5.0h~\mathrm{Mpc}^{-1}$ because the resolution of our
simulations is not high enough, but this could be overcome in
future higher-resolution simulations.

To show that our ray-tracing simulations produce reasonable
results, we first consider the convergence and shear maps from a
chosen realisation of tiling solution, and these are shown in
Fig.~\ref{fig:map}. We have computed the convergence field
$\kappa(\vec{\xi})$, using the two methods outlined in
Sections~\ref{subsect: method1} (Method A,
Eq.~(\ref{eq:int_by_part}), left panel of Fig.~\ref{fig:map}) and
\ref{subsect: method2} (Method B, Eq.~(\ref{eq:exact_kappa}),
right panel of Fig.~\ref{fig:map}). The two methods give almost
identical results, and as we have checked, the difference is in
general well within the percent level. The shear field
$\left(\gamma_1,\gamma_2\right)$ is also calculated using two
methods: method A using an equivalence of
Eq.~(\ref{eq:exact_kappa}) as described in the figure caption
(left panel), and method B which is often used in the literature,
namely by Fourier transforms of the convergence field which only
works in the weak-lensing regime (right panel). The shear fields
are shown in rods along with the convergence map shown as images.
Again, the agreement for the shear field is very good, indicating
that our ray-tracing algorithm works well.

We also show the lensing convergence power spectrum measured from
our ray-tracing simulations in Fig.~\ref{fig:pkappa}. Due to our
limit field of view of $(5^{\circ}\times5^{\circ})$, we cannot
measure the spectrum at multiple moment $\ell<100$. Also, there is
a rolloff of power at $\ell>2000$, which is because (1) the
resolution for our $N$-body simulations is not high enough and (2)
the {\tt TSC} density-assignment scheme smooths out the
small-scale structure more than the {\tt CIC} and {\tt NGP}
schemes do. Both factors tend to suppress the convergence power
spectrum at high $\ell$ and we hope to solve this problem by using
higher resolution simulations and more suitable interpolation
schemes, which is left for our future study. Otherwise, we find
that the ray-tracing result agrees reasonably well with the
analytic prediction using the fitting formula for matter power
spectrum by \citet{setal2003} in some $\ell$ range, \ie,
$100<\ell<2000$. On some scales, we see that the numerical result
is slightly higher than the theoretical prediction. This is
however as expected because we have seen from Fig.~\ref{fig:power}
that the $N$-body simulations give a higher matter power spectrum
than the \citet{setal2003} fit on some scales. The fact that a
difference in the matter power spectra from simulations and
analytic fitting could cause differences in the computed
convergence power spectra has been reported and discussed by many
authors, {\it e.g.}, \citet{vw2003, hhws, prts2010}. In
Fig.~\ref{fig:pkappa}, we overplot the expected observational
uncertainty from DES using the survey parameters $f_{\rm
sky}\sim0.12, \bar{n}_{\rm g}=10$/arcmin$^2$, $\gamma_{\rm
int}^2\sim0.16$ where $f_{\rm sky}, \bar{n}_{\rm g}$ and
$\gamma_{\rm int}^2$ denote the sky coverage, number of galaxies
per arc-minute squared and the mean-square intrinsic ellipticity,
respectively.

Note that in our numerical simulations we have not included the
lens-lens coupling and second-order corrections to the Born
approximation. In Appendix~\ref{appen:A} we will outline how these
can be incorporated in future higher-resolution simulations.

\section{Discussion and Conclusion}

\label{sect:con}

The correlations in the distorted images of distant galaxies,
induced by cosmic shear, hold information about the distribution
of matter on a wide range of scales in the universe. In order to
take full advantage of current and future weak lensing data sets
to constrain cosmology, using information from both the linear and
non-linear regimes, one needs a sophisticated algorithm to measure
the shear and convergence fields from $N$-body simulations, and to
construct statistical quantities. This is traditionally done using
the `plane-by-plane' discrete lens-plane algorithm -- trace the
virtual light rays and calculate the deflection caused by the
density field projected onto a number of 2-D lensing planes.

In this work, we propose an improved ray-tracing algorithm. We
calculate the deflection of the light rays caused by the detailed
3-D density fields living on the natural simulation mesh, rather
than the simplified density distribution projected onto some 2-D
planes. We evaluate the shear and convergence fields by
analytically integrating the deflection as the light rays go
through the individual simulation grid cells. This approach is
easy to implement and  computationally inexpensive. It avoids
numerical integration, and expensive data storage since it is
performed on the fly. We apply the algorithm to our simulations,
and find good agreement with the \citet{setal2003} fit, and
consistency with the published results in \citet{shttym2009}.

The on-the-fly l.o.s.~integration is computationally economic. In the {\tt RATANA} code, most
computation time is spent on the $N$-body part. Suppose $N_d^3$ is the number of grid cells
in our mesh, then the {\tt FFT} requires $3N_d^3\log_2N_d$ operations each time step, not including
other operations such as differencing the potential to obtain the force on the mesh, assigning particles
and computing densities on all the grid cells and particle movements. In contrast, if we let 
$N_{\rm{los}}=N_d$ (which is enough for accuracy), then there are only $N_d^2$ rays to trace, and 
for each ray we have $\leq10^2$ operations. We have checked the simulation log file and found that
there is no significant difference in the times used by each step before and after the ray-tracing part 
of {\tt RATANA} has been triggered. 

Analytic formulae are often more useful than purely numerical results in 
tracing the physical contents of a theory. For example, in Eqs.~(\ref{eq:los_int}, 
\ref{eq:ds}), it is easy to check which terms contribute the most to the final result: 
obviously, in the small-angle limit, {\it i.e.}, $\theta,\psi\ll1$, terms involving $d_3, d_4$, 
and a large part of $d_2$ could be neglected because $\sin\theta, \sin\psi\ll1$; also 
at least one of $a,b,c$ vanishes and $abc=0$ for all grid cells, further simplifying $d_1, d_2$; furthermore, 
terms in Eq.~(\ref{eq:los_int}) with coefficient $1/\chi_s$ contribute little because 
$\chi_s\gg\tilde{\chi}_{u,l}$. Such observations can be helpful in determining which
terms have important effects in certain regimes.

Note that the dependence on $\chi_s$ [cf.~Eq.~(\ref{eq:los_int})] could be taken out 
of the analytical integration, meaning that the algorithm can be straightforwardly generalised
to include multiple source redshifts with very little extra computational effort (mainly 
in determining where to start  the integration for a given source redshift). The algorithm can
also be easily generalised to compute the flexion, which depends on higher-order derivatives of 
the lensing potential, and is expected to give more accurate results than the multiple-lens-plane
approximation. 

The algorithm has many other flexibilities too. As an example, the
analytic integration of the projected density and potential
fields along the l.o.s.~can be performed on an adaptive rather
than a regular grid with careful programming, which means that
higher resolution can be achieved in high density regions, as in
the adaptive {\tt PM} simulations. Also, the analytic
integration can be easily generalised to other algorithms to
compute the 3-D shear field \citep{cbt1999}.

We also give prescriptions to include second-order corrections to
the results, such as the lens-lens coupling and corrections to the
Born approximation, in Appendix~\ref{appen:A}. It is interesting
to note that, by running the $N$-body simulations backwards in
time, we can still compute the convergence and shear fields
on-the-fly even if the light rays are not straight.

To conclude, the algorithm described here is efficient and
accurate, and is suitable for the future ray-tracing simulations
using very large $N$-body simulations. It will be interesting to
apply it to study the higher-order statistics of the shear field
and the lensing excursion angles, and these will be left for
future work.

\section*{Acknowledgments}

The work described here has been performed under the {\tt
HPC-EUROPA} project, with the support of the European Community
Research Infrastructure Action under the {\tt FP8} "Structuring
the European Research Area" Programme. The $N$-body simulations
are performed on the {\tt SARA} supercomputer in the Netherlands,
and the post-precessing of data is performed on {\tt COSMOS}, the
UK National Cosmology Supercomputer. The \citet{setal2003} fit
results for the matter and convergence power spectra are computed
using the {\tt CAMB} code. The nonlinear matter power spectrum is
measured using {\tt POWMES} \citep{cjnp2009}. We would like to
thank Henk Hoekstra for being the local host for the {\tt HPC-EUROPA} project, 
and Henk Hoekstra, David Bacon, Kazuya Koyama for useful discussions. BL is
supported by Queens' College at University of Cambridge and STFC rolling grant in DAMTP,
LK is supported by the Royal Society, GBZ is supported by STFC grant ST/H002774/1.

\onecolumn
\appendix

\section{Beyond the First-order Approximations}

\label{appen:A}

In the attempt to trace light rays on the fly, we set up a
bundle of l.o.s. before the $N$-body simulation starts. But
because we do not know the exact paths of those light rays which
finally end up at the observer, we have to assume that they are
straight lines even though they are not in reality. This so-called
Born approximation is generally quite good in the weak lensing
regime, but can lead to non-negligible errors on small scales
\citep{hhws}. Furthermore, in the above treatment we have also
neglected the lens-lens coupling, which accounts for the fact that
the lenses themselves (the large-scale structure) are distorted by
the lower-redshift matter distribution.

\citet{hhws} take account of the lens-lens coupling and
corrections to the Born approximation using the
multiple-lens-plane approximation. In such an approach, the light
rays get deflected and their paths are recomputed when and only
when they pass by a discrete lens plane.

Since our algorithm goes beyond the discrete lens-plane
approximation and is able to trace the detailed matter
distribution, we want to generalise it to include those
corrections as well. In this Appendix we shall derive an
analytical formula for the distortion matrix with the lens-lens
coupling taken into account, and describe how the corrections to
the Born approximation can be incorporated as well.

Obviously, to go beyond the Born approximation, the light rays are
no longer straight and thus the l.o.s~cannot be set up before the
$N$-body simulation has finished. Instead, we have to start from
the observer today and go backwards in time to compute the
distortion matrix Eq.~(\ref{eq:distortmat}). We shall discuss
below how this could be realised in practice, but at this moment
let us simply assume that we can go backwards in time, and know
the value of the lensing potential $\Phi$ and its derivatives
along the l.o.s..

\subsection{Corrections to the Born Approximation}

The corrections to the Born approximation are easy to implement.
According to Eq.~(\ref{eq:xi}), the total deflection of a light
ray is the sum of the deflections by the matter in each grid that
ray passes on its way towards the lensing source. Suppose
$\vec{\xi}^{(n)}$ denotes the value of $\vec{\xi}$ after the light
ray crosses the $n$-th grid on its way ($n$ increases with the
distance from the observer, $n=1$ corresponds to the grid which
the observer is in, and $\vec{\xi}^{(0)}=\vec{\xi}_0$), then
\begin{eqnarray}\label{eq:borncorrect}
\vec{\xi}^{(n)} &=& \vec{\xi}^{(n-1)}
-2\int^{\chi^{(n)}_u}_{\chi^{(n)}_l}\frac{\chi_s-\chi}{\chi\chi_s}\vec{\nabla}_{\vec{\xi}}\Phi
d\chi,
\end{eqnarray}
where
$\chi^{(n)}_u\equiv\min\left\{\chi^{\mathrm{ts}}_{u},\chi^{(n)}_B\right\}$
and
$\chi^{(n)}_l\equiv\max\left\{\chi^{\mathrm{ts}}_{l},\chi^{(n)}_A\right\}$,
in which $\chi^{\mathrm{ts}}_{u}>\chi^{\mathrm{ts}}_l$ are
respectively the $\chi$-values at the two ends of the current time
step, and $\chi^{(n)}_B>\chi^{(n)}_A$ the $\chi$-values of the two
intersections between the light ray and the $n$-th grid. Using the
expressions given in Sect.~\ref{subsect: method2}, it is easy to
write $\nabla_{\xi_1}\Phi$ and $\nabla_{\xi_2}\Phi$ in terms of
polynomials of $\chi$. Then the above integral can be performed
analytically as before. In this way, each time the light ray
crosses a grid, we update its orientation according to the above
equation, and thus the corrections to the Born approximation can
be incorporated.

Note that in this approach the light rays are deflected many more
times than in the multiple-lens-plane approximation and the
detailed matter distribution has been \tc {fully taken} account of.

\subsection{Lens-lens Coupling}

As mentioned earlier, the lens-lens coupling has been neglected in
the above treatment because in Eq.~(\ref{eq:distortmat}) we have
used the approximation
$\nabla_{\chi_{0i}}\approx\nabla_{\chi_{i}}$. Let us now have a
look at what happens when this approximation is dropped.

Note that in the expression
\begin{eqnarray}
A_{ij} &\equiv&
-2\int^{\chi_s}_0\frac{g(\chi,\chi_s)}{\chi^2}\nabla_{\xi_{0i}}\nabla_{\xi_j}\Phi\left(\chi,\vec{\xi}\right)
d\chi + \delta_{ij},
\end{eqnarray}
the argument of $\Phi$ is $\vec{\xi}$ while one of the derivatives
is with respect to $\vec{\xi}_0$. We can utilise the chain rule
to write
$\nabla_{\xi_{0i}}=\left(\nabla_{\xi_{0i}}\xi_j\right)\nabla_{\xi_{j}}=A_{ij}\nabla_{\xi_{j}}$
where we have used the definition of $A_{ij}$ given in
Sect.~\ref{subsect:oldmethod}. Then the above equation becomes
\begin{eqnarray}\label{eq:lenslens}
A^i_{\ j}\left(\chi_s,\vec{\xi}\right)
&\equiv &\delta^i_{\
j}-2\int^{\chi_s}_0g(\chi,\chi_s)\nabla^i\nabla_k\Phi\left(\chi,\vec{\xi}\right)A^k_{\
j}\left(\chi,\vec{\xi}\right)d\chi,
\end{eqnarray}
where for simplicity we have used $\nabla_i=\nabla_{\xi_i}$. With
the $A^k_{\ j}$ term in the integrand, Eq.~(\ref{eq:lenslens}) now
includes the lens-lens coupling, and will be our starting point
here.

Again, let us consider the integral in Eq.~(\ref{eq:lenslens})
after the light ray crosses the $n$-th grid on its way towards the
lensing source. The discrete version of Eq.~(\ref{eq:lenslens}) is
\begin{eqnarray}\label{eq:lenslens2}
^{(n)}A^i_{\ j}
&=& ^{(n-1)}A^i_{\ j} -
2\int^{\chi^{(n)}_u}_{\chi^{(n)}_l}\frac{\chi\left(\chi_s-\chi\right)}{\chi_s}~^{(n)}A^k_{\
j} \nabla^i\nabla_k\Phi d\chi
\end{eqnarray}
where $^{(n)}A^i_{\ j}$ is the value of $A^i_{\ j}$ after the
light ray has crossed the $n$-th grid, and $^{(0)}A^i_{\
j}=\delta^i_{\ j}$ as is easy to see. This formula has three
advantages as compared to the multiple-lens-plane approximation:
\begin{enumerate}
    \item As before, the light rays between $z=0$ and $z_s$ are
    divided into many more segments, and the fine structure of the matter distribution
    is included naturally, without squeezing the matter and using
    impulse approximations.
    \item As will be shown below, the integration can be evaluated
    analytically rather than numerically.
    \item Note that we can use $^{(n)}A^k_{\ j}$ rather than $^{(n-1)}A^k_{\
    j}$ in the integrand, which will give more accurate results,
    because using $^{(n-1)}A^k_{\ j}$ would mean that the
    contribution to the lens-lens coupling from the matter in the
    $n$-th grid is ignored. In the multiple-lens-plane
    approximation which typically uses $20\sim30$ lens planes,
    the $n$-th plane could contain a significant amount of matter,
    and neglecting its contribution could make the results less
    accurate.
\end{enumerate}

Eq.~(\ref{eq:lenslens2}) is exact, but we only want the result to
second order in $\nabla\nabla\Phi$. Therefore we can iterate once
and write an approximate solution as
\begin{eqnarray}\label{eq:lenslens3}
^{(n)}A^i_{\ j}
&\approx& ^{(n-1)}A^i_{\ j} - 2~^{(n-1)}A^k_{\
j}\int^{\chi^{(n)}_u}_{\chi^{(n)}_l}d\chi
\frac{\chi\left(\chi_s-\chi\right)}{\chi_s}\nabla^i\nabla_k\Phi\\\nonumber
&&+4\int^{\chi^{(n)}_u}_{\chi^{(n)}_l}d\chi\frac{\chi\left(\chi_s-\chi\right)}{\chi_s}\nabla^i\nabla_k\Phi\left(\chi,\vec{\xi}\right)
\int^{\chi}_{\chi^{(n)}_l}d\chi'\frac{\chi'\left(\chi-\chi'\right)}{\chi}\nabla^k\nabla_j\Phi\left(\chi',\vec{\xi}\right).\nonumber
\end{eqnarray}
Following the approach taken in Sect.~\ref{subsect: method1} we
can write
\begin{eqnarray}\label{eq:use}
\nabla^i\nabla_j\Phi\left(\chi,\vec{\xi}\right) &=& \sum_{N=1}^{4}
\left(\mathbf{K}_{N}\right)^{i}_{\
j}\left(\chi-\chi_A^{(n)}\right)^{N-1},
\end{eqnarray}
where $\chi_A^{(n)}$ is defined in Eq.~(\ref{eq:borncorrect}), and
$\mathbf{K}_N$ ($N\in\{1,2,3,4\}$) is a $2\times2$ matrix whose
$ij$-component depends on the orientation of the l.o.s.~segment
inside the $n$-th grid (where it is taken to be straight) and the
values of $\nabla^i\nabla_j\Phi$ at the vertices of the $n$-th
grid. Note however that $\mathbf{K}_N$ is independent of $\chi$.
The expressions are similar to the $d_N$s defined in
Sect.~\ref{subsect: method1} and we shall not write them
explicitly here.

Substituting Eq.~(\ref{eq:use}) into Eq.~(\ref{eq:lenslens3}), we
find
\begin{eqnarray}\label{eq:lenslens4}
^{(n)}A^i_{\ j} &=&~^{(n-1)}A^i_{\ j} -
2\sum^{4}_{N=1}I_1(N)\left(\mathbf{K}_{N}\right)^{i}_{\
k}~^{(n-1)}A^k_{\ j} + 4\sum^{4}_{N=1}\sum^{4}_{M=1}I_2(N,M)\left(\mathbf{K}_{N}\right)^{i}_{\
k}\left(\mathbf{K}_{M}\right)^{k}_{\ j}
\end{eqnarray}
in which we have written (again, by defining
$\tilde{\chi}\equiv\chi-\chi_A^{(n)}$,
$\tilde{\chi}'\equiv\chi'-\chi_A^{(n)}$ and
$\tilde{\chi}^{(n)}_{u,l}\equiv\chi^{(n)}_{u,l}-\chi_A^{(n)}$)
\begin{eqnarray}
I_1(N) &\equiv&
\int^{\chi^{(n)}_u}_{\chi^{(n)}_l}d\chi\frac{\chi\left(\chi_s-\chi\right)}{\chi_s}
\left(\chi-\chi_A^{(n)}\right)^{N-1}\nonumber\\
&=&
\frac{1}{N}\chi_A^{(n)}\left(1-\frac{\chi^{(n)}_A}{\chi_s}\right)\left[\left(\tilde{\chi}^{(n)}_{u}\right)^{N}
- \left(\tilde{\chi}^{(n)}_{l}\right)^{N}\right]
+\frac{1}{N+1}\left(1-\frac{2\chi^{(n)}_A}{\chi_s}\right)\left[\left(\tilde{\chi}^{(n)}_{u}\right)^{N+1}
- \left(\tilde{\chi}^{(n)}_{l}\right)^{N+1}\right]\nonumber\\
&&-\frac{1}{N+2}\frac{1}{\chi_s}\left[\left(\tilde{\chi}^{(n)}_{u}\right)^{N+2}
- \left(\tilde{\chi}^{(n)}_{l}\right)^{N+2}\right],
\end{eqnarray}
and
\begin{eqnarray}
I_2(N,M) &\equiv&
\int^{\chi^{(n)}_u}_{\chi^{(n)}_l}d\chi\frac{\chi_s-\chi}{\chi_s}\left(\chi-\chi_A^{(n)}\right)^{N-1}
\int^{\chi}_{\chi^{(n)}_l}d\chi'\chi'(\chi-\chi')\left(\chi'-\chi_A^{(n)}\right)^{M-1}\nonumber\\
&=& -\frac{\left(\tilde{\chi}^{(n)}_{u}\right)^{M+N+3} -
\left(\tilde{\chi}^{(n)}_{l}\right)^{M+N+3}}{(M+1)(M+2)(M+N+3)}\frac{1}{\chi_s}
+\frac{\left(\tilde{\chi}^{(n)}_{u}\right)^{M+N+2} -
\left(\tilde{\chi}^{(n)}_{l}\right)^{M+N+2}}{(M+1)(M+2)(M+N+2)}\left(1-\frac{\chi_A^{(n)}}{\chi_s}\right)\nonumber\\
&&-\frac{\left(\tilde{\chi}^{(n)}_{u}\right)^{M+N+2} -
\left(\tilde{\chi}^{(n)}_{l}\right)^{M+N+2}}{M(M+1)(M+N+2)}\frac{\chi_A^{(n)}}{\chi_s}
+\frac{\left(\tilde{\chi}^{(n)}_{u}\right)^{M+N+1} -
\left(\tilde{\chi}^{(n)}_{l}\right)^{M+N+1}}{M(M+1)(M+N+1)}\chi_A^{(n)}\left(1-\frac{\chi_A^{(n)}}{\chi_s}\right)\nonumber\\
&&+\left[\frac{1}{M}\frac{\chi_A^{(n)}}{\chi_s}+\frac{1}{M+1}\frac{\chi_l^{(n)}}{\chi_s}\right]\left(\chi_l^{(n)}\right)^{M}
\frac{1}{N+2}\left[\left(\tilde{\chi}^{(n)}_{u}\right)^{N+2} -
\left(\tilde{\chi}^{(n)}_{l}\right)^{N+2}\right]\nonumber\\
&&-\left(1-\frac{\chi_A^{(n)}}{\chi_s}\right)
\left[\frac{1}{M}\chi_A^{(n)}+\frac{1}{M+1}\chi_l^{(n)}\right]\left(\chi_l^{(n)}\right)^{M}
\frac{1}{N+1}\left[\left(\tilde{\chi}^{(n)}_{u}\right)^{N+1} -
\left(\tilde{\chi}^{(n)}_{l}\right)^{N+1}\right]\nonumber\\
&&-\left[\frac{1}{M+1}\frac{\chi_A^{(n)}}{\chi_s}+\frac{1}{M+2}\frac{\chi_l^{(n)}}{\chi_s}\right]\left(\chi_l^{(n)}\right)^{M+1}
\frac{1}{N+1}\left[\left(\tilde{\chi}^{(n)}_{u}\right)^{N+1} -
\left(\tilde{\chi}^{(n)}_{l}\right)^{N+1}\right]\nonumber\\
&&+\left(1-\frac{\chi_A^{(n)}}{\chi_s}\right)
\left[\frac{1}{M+1}\chi_A^{(n)}+\frac{1}{M+2}\chi_l^{(n)}\right]\left(\chi_l^{(n)}\right)^{M+1}
\frac{1}{N}\left[\left(\tilde{\chi}^{(n)}_{u}\right)^{N} -
\left(\tilde{\chi}^{(n)}_{l}\right)^{N}\right].
\end{eqnarray}

The above expressions look rather heavy, however, they are
analytic and as a result are very easy to implement in the
ray-tracing simulation codes, by writing functions that take
$M,N,\chi_A^{(n)},\chi_u^{(n)},\chi_l^{(n)}$ as parameters and
return $I_1, I_2$ as outputs. Furthermore, since the grid size
($<0.2h^{-1}$~Mpc) in the $N$-body simulations is small enough
compared with the typical inter-plane distances in the
multiple-lens-plane approximations ($10\sim100h^{-1}$~Mpc), we can
drop the $I_2(N,M)$ terms to a very good approximation, which will
greatly simplify the results.

Note that the distortion matrix $A^i_{\ j}$ computed in this way
is not symmetric because of the matrix multiplications. However,
using Eq.~(\ref{eq:distortmat2}), it is straightforward to compute
$\gamma_2=-\left(A^1_{\ 2}+A^2_{\ 1}\right)/2$. In addition, we
could also calculate the rotation $\omega$ as $\omega=\left(A^2_{\
1}-A^1_{\ 2}\right)/2$

\subsection{Going Back In Time}

As mentioned above, to include the actual deflections of the light
rays which end up at the observer, we have to start from the
observer and go backwards in time until encountering the source.
This obviously can only be done after the $N$-body simulation has
finished.

One way to go backwards is to record the information about the
gravitational potential $\Phi$ and its derivatives in a light cone
during the simulation, and then post-process the light-cone data.
This means that a large amount of dump data has to be stored.

Alternatively, one can think of running the $N$-body simulation
"backwards". To be more explicit, the simulation is first run in
the forward direction from a high redshift until today, and we
obtain the particle positions and velocities at present; then we
reverse the directions of the gravitational force and the particle
velocities, and evolve the system back until $z_s$ using the same
time-stepping scheme as in the forward simulation. In this way,
the actual light rays and distortion matrix could be built up on
the fly, and there is no need to store a lot of dump data.

\label{lastpage}

\end{document}